\documentclass[aps,prd,preprintnumbers,nofootinbib,11pt]{revtex4}
\usepackage{graphicx}
\usepackage{amsmath,amssymb,bm}

\begin{document}

\title{Exact Hairy Black Holes}
\thanks{ Prepared for the Proceedings of \textit{Relativity and Gravitation:
100 years after Einstein in Prague}, Prague, 25-29 June, 2012.}
\author{Andr\'{e}s Anabal\'{o}n}
\email{andres.anabalon@uai.cl}

\address{Departamento de Ciencias, Facultad de Artes
Liberales, Facultad de Ingenier\'{\i}a y Ciencias, Universidad
Adolfo Ib\'{a}\~{n}ez, Vi\~{n}a Del Mar, Chile.}

\begin{abstract}
This proceeding reviews the recent finding of a certain class of, regular on
and outside the horizon, exact hairy black hole solutions in four
dimensional general relativity. Their construction follow from the
integrability of a cohomogeneity two Weyl rescaling of the Carter-Debever
ansatz in the presence of an arbitrary number of scalar fields with an
arbitrary self interaction and an arbitrary non-minimal coupling to the
scalar curvature. Two field equations, independent of the specific form of
the energy momentum tensor, are used to integrate the metric. The remaining
ones fix the form of the scalar field self interaction. The cohomogeneity
one black holes are described and are shown to encompass all the exact,
regular in the domain of outer communications, uncharged, black holes with a
minimally coupled scalar hair, available in the literature.
\end{abstract}

\maketitle

%% for revtex4

\section{Introduction and Discussion}

The field of exact solutions in gravity, as well as its interpretation, is
as old as general relativity and the research group at Charles university,
and their collaborators, are well known for their contributions to this
subject. Many of them can be found in the review \cite{Bicak:2000ea} or the
book \cite{GP}. From the black hole uniqueness theorems it is already well
known that at least in four dimensions, the asymptotically flat, stationary
and regular black holes in the electrovac case are exhausted, for references
see \cite{Chrusciel:2012jk}. Therefore, is natural to attempt to extend
these studies when other matter fields are included. Indeed, the study of
the minimally coupled scalar field have a prominent role in the construction
of black holes. In the static, asymptotically flat case, the minimally
coupled no-hair conjecture was shown to be true for convex potentials \cite%
{Bekenstein:1972ny}, and, more generally, for potentials satisfying the
strong \cite{Heusler:1992ss} and weak energy condition \cite{Sudarsky:1995zg}%
. These studies have their counterpart in Brans-Dicke \cite{Hawking:1972qk}
and more generally in scalar-tensor theories \cite{Sotiriou:2011dz}, showing
that whenever the scalar field potential satisfy the weak energy condition
in the Einstein frame and the black hole spacetime is stationary and
asymptotically flat it must be Kerr. When the scalar field satisfy the null
energy condition an exact family of spherically symmetric black hole
solution has been recently constructed \cite{Anabalon:2012ih}.

When the cosmological constant is negative exact, uncharged, $AdS_{4}$ hairy
black hole solutions has been extensively studied \cite{Martinez:2004nb,
Kolyvaris:2009pc, Anabalon:2012sn, Toldo:2012ec, Anabalon:2012ta}. There is
a precise conjecture on the non-existence of spherically symmetric black
holes in AdS for scalar field potentials that comes from \textquotedblleft
the right\textquotedblright\ superpotential \cite{Hertog:2006rr}. These
solutions are interesting at the light of the AdS/CFT conjecture. In
particular, in four dimensions, and when the scalar field is charged, they
define the setting for the AdS/Condensed matter correspondence \cite%
{Hartnoll:2008vx}. When the cosmological constant is positive the black
holes have also attracted some attention of the community \cite%
{Zloshchastiev:2004ny}.

This proceeding intends to shortly summarize my recent contributions to the
subject. I have followed the idea that stationary and axisymmetric
spacetimes that have a hidden symmetry, in the form of a conformal Killing
tensor, should allow for a complete integrability of some form of
non-trivial self interaction of the scalar field. Therefore, in \cite%
{Anabalon:2012ta} I explicitly showed that, starting with the ansatz that
contains all the vacuum Petrov type D solutions, it is possible to integrate
the system in the presence of a non-minimally coupled scalar field or a
non-linear sigma model. It is very interesting to note that the self
interaction of the scalar field is completely fixed by the form of the
metric ansatz and, therefore, the scalar field potential is an output of the
analysis. While these results are presented in the Einstein frame, their
extension to a Scalar-Tensor theories in some Jordan Frame or $F(R)$ theory
is straightforward.

The scalar field potential turns out to contain as special cases all the
exact hairy (A)dS black holes available in the literature. The static
solutions are black holes, continuosly connected with the Schwarzschild
(A)dS solution, and can be generalized to include non-minimally coupled
gauge fields \cite{Salvio:2012at}. For the asymptotically AdS black holes,
with cosmological constant $\Lambda =-3/l^{2}$, the scalar field mass is $%
m^{2}=-2/l^{2}$, which is above the Breitenlohner-Freedman bound, $%
m^{2}=-9/4l^{2}$, ensuring the perturbative stability of these black holes.
This mass is the one of the scalar fields of the $U(1)^{4}$ truncation of
gauged $N=8$ supergravity \cite{Duff:1999gh} and a sub class of the
solutions can be embedded in this supergravity theory.

The outlook of the proceeding is as follows. In the second section the
general integrability of the ansatz with two killing vectors is reviewed and
in the third section the static case and its special limits is presented.

\section{The integrable system with two Killing vectors.}

The conventions of the proceeding are given by the action principle

\begin{equation}
S(g,\phi )=\int d^{4}x\sqrt{-g}\left[ \frac{R}{2\kappa }-\frac{1}{2}g^{\mu
\nu }\partial _{\mu }\phi \partial _{\nu }\phi -\frac{\xi }{12}\phi
^{2}R-V(\phi )\right] ,  \label{AP}
\end{equation}%
where $\kappa =8\pi G$. We are interested in studying a cohomogeneity two
Weyl rescaling of the Carter-Debever ansatz \cite{Carter:1968ks, Debever},
also studied by Plebanski \cite{Pleba}:

\begin{eqnarray}
ds^{2} &=&S(q,p)\left( \frac{1+p^{2}q^{2}}{Y(q)}dq^{2}+\frac{1+p^{2}q^{2}}{%
X(p)}dp^{2}-\frac{Y(q)}{1+p^{2}q^{2}}\left( p^{2}d\tau +d\sigma \right)
^{2}\right)  \notag \\
&&+S(q,p)\frac{X(p)}{1+p^{2}q^{2}}\left( d\tau -q^{2}d\sigma \right) ^{2}
\end{eqnarray}

When $S(q,p)=p^{-2},$ this metric contains the Kerr-Newman hole with a
cosmological constant. Letting $S(q,p)$ free, this metric can be integrated
in vacuum, and with the same Maxwell field than in the Kerr-Newman case, the
Plebanski-Demianski spacetime arises \cite{Plebanski:1976gy}.

The observation made in \cite{Anabalon:2012ta}, is that for stationary and
axisymmetric scalar fields, $\phi =\phi (q,p)$, the energy momentum tensor
of a minimally coupled scalar field 
\begin{equation}
T_{\mu \nu }=\partial _{\mu }\phi \partial _{\nu }\phi -\frac{1}{2}g_{\mu
\nu }\left( \partial \phi \right) ^{2}-g_{\mu \nu }V(\phi )
\end{equation}%
is such that $T_{\sigma }^{\tau }=0=T_{\tau }^{\sigma }$ and therefore the
Einstein equations, $R_{\nu }^{\mu }-\frac{1}{2}\delta _{\nu }^{\mu
}R=\kappa T_{\nu }^{\mu }$ imply $R_{\sigma }^{\tau }=0=R_{\tau }^{\sigma }$%
. These two equations are enough to completely the metric functions and the
solution is

\begin{eqnarray}
X(p) &=&C_{0}+C_{2}p^{2}+C_{4}p^{4}+C_{1}p^{-\nu +2}+C_{3}B_{3}p^{\nu +2},
\label{1} \\
Y(q) &=&C_{4}-C_{2}q^{2}+C_{0}q^{4}+C_{3}C_{1}q^{-\nu +2}+B_{3}q^{\nu +2},
\label{2}
\end{eqnarray}%
\begin{equation}
S(q,p)=C\frac{p^{\nu -1}q^{\nu -1}}{(C_{3}p^{\nu }+q^{\nu })^{2}}.  \label{3}
\end{equation}

This solution reduces to the Plebanski-Demianski spacetime when $\nu =\pm 1$%
. The remaining Einstein equations fix the scalar field and the scalar field
potential to a very precise form. The same process can be done when the
scalar field is non-minimally coupled to gravity and, more generally, when a
non-linear sigma model is the source of the Einstein equations.

To extract more physical information let us study the cohomogeneity one
black holes.

\section{The static black holes.}

The static limit of the previous configuration is

\begin{equation}
ds^{2}=\Omega (r)(-F(r)dt^{2}+\frac{dr^{2}}{F(r)}+d\Sigma ^{2}),  \label{M1}
\end{equation}%
\begin{equation}
\Omega (r)=\frac{\nu ^{2}\eta ^{\nu -1}r^{\nu -1}}{\left( r^{\nu }-\eta
^{\nu }\right) ^{2}},\qquad \phi =l_{\nu }^{-1}\ln (r\eta ^{-1}),
\end{equation}

\begin{equation}
F(r)=\frac{r^{2-\nu }\eta ^{-\nu }\left( r^{\nu }-\eta ^{\nu }\right) ^{2}k}{%
\nu ^{2}}+\left( \frac{1}{\left( \nu ^{2}-4\right) }-\left( 1+\frac{\eta
^{\nu }r^{-\nu }}{\nu -2}-\frac{\eta ^{-\nu }r^{\nu }}{\nu +2}\right) \frac{%
r^{2}}{\eta ^{2}\nu ^{2}}\right) \alpha -\frac{\Lambda }{3},  \label{M3}
\end{equation}%
where $l_{\nu }=\left( \frac{2\kappa }{\nu ^{2}-1}\right) ^{\frac{1}{2}}$
and $d\Sigma ^{2}$ is the line element of a surface of constant curvature $k$%
. $\eta $ is the only integration constant of the black hole. The solution
and theory are invariant under the transformation $\nu \rightarrow -\nu $.

The scalar field potential is

\begin{align}
V(\phi )& =\frac{\Lambda \left( \nu ^{2}-4\right) }{6\kappa \nu ^{2}}\left( 
\frac{\nu -1}{\nu +2}e^{-\left( \nu +1\right) \phi l_{\nu }}+\frac{\nu +1}{%
\nu -2}e^{\left( \nu -1\right) \phi l_{\nu }}+4\frac{\nu ^{2}-1}{\nu ^{2}-4}%
e^{-\phi l_{\nu }}\right) +  \notag \\
& \frac{\alpha }{\nu ^{2}\kappa }\left( \frac{\nu -1}{\nu +2}\sinh (\left(
1+\nu \right) \phi l_{\nu })+\frac{\nu +1}{\nu -2}\sinh (\left( 1-\nu
\right) \phi l_{\nu })+4\frac{\nu ^{2}-1}{\nu ^{2}-4}\sinh \left( \phi
l_{\nu }\right) \right) .  \label{pot1}
\end{align}

It is easy to see from the form of the metric, and without any reference to
the details of the solution itself, that it is possible to introduce
Eddington-Finkelstein coordinates $u_{\mp }=t\pm \int \frac{dr}{F(r)},\,$%
which allows to cover either the black hole ($u_{-}$) or the white hole ($%
u_{+}$). The asymptotically flat solution has a single horizon from which it
follows that the Penrose diagram is the same than the Schwarzschild black
hole.

The energy momentum of the scalar field, in a comoving tetrad, has the form $%
T^{ab}=diag(\rho ,p_{1},p_{2},p_{2})$ and, in the static regions of the
spacetime, defined by $F(r)>0$, satisfies the null energy condition 
\begin{equation}
\rho +p_{2}=0,\qquad \rho +p_{1}=\frac{\left( \nu ^{2}-1\right) \left(
r^{\nu }-\eta ^{\nu }\right) ^{2}F(r)}{2r\nu ^{2}\eta ^{\nu -1}r^{\nu }}>0.
\end{equation}%
In the hairless limit, $\nu =1$, the change of coordinates $r=\eta -\frac{1}{%
y}$ brings the hairy solution (\ref{M1})-(\ref{M3}) to the familiar
Schwarzschild de Sitter black hole

\begin{equation}
ds^{2}=-(k-\frac{2M}{y}-\frac{\Lambda }{3}y^{2})dt^{2}+\frac{dy^{2}}{k-\frac{%
2M}{y}-\frac{\Lambda }{3}y^{2}}+y^{2}d\Sigma .  \label{K}
\end{equation}%
where $M=\frac{3\eta ^{2}k+\alpha }{6\eta ^{3}}$.

The parameterization of the black holes has been chosen such that its
leading order at $r=\eta $ is either Minkwoski, anti de Sitter or de Sitter
in the following form

\begin{equation}
ds_{r=\eta }^{2}=\frac{1}{(r-\eta )^{2}}\left( -(k(r-\eta )^{2}-\frac{%
\Lambda }{3})+\frac{dr^{2}}{(k(r-\eta )^{2}-\frac{\Lambda }{3})}+d\Sigma
^{2}\right) .
\end{equation}%
The easiest way to see that there is always an $\alpha $ such that $F(r)$
has a simple zero is to see that the equation $F(r_{+})=0$ is linear in $%
\alpha $

\begin{equation}
0=\frac{r_{+}^{2-\nu }\eta ^{-\nu }\left( r_{+}^{\nu }-\eta ^{\nu }\right)
^{2}k}{\nu ^{2}}+\left( \frac{1}{\left( \nu ^{2}-4\right) }-\left( 1+\frac{%
\eta ^{\nu }r_{+}^{-\nu }}{\nu -2}-\frac{\eta ^{-\nu }r_{+}^{\nu }}{\nu +2}%
\right) \frac{r_{+}^{2}}{\eta ^{2}\nu ^{2}}\right) \alpha -\frac{\Lambda }{3}%
,
\end{equation}%
therefore is possible to solve this equation for $\alpha $ for any value of
the other parameters.

As a final remark it is instructive to compare the behavior of these
solutions in $AdS$, with the asymptotic form given in \cite{Henneaux:2006hk}%
. When the backreaction is ignored, an scalar field with mass $m$ minimally
coupled to an $AdS$ background has the well known fall off $\phi \sim \frac{a%
}{\rho ^{\Delta _{-}}}+\frac{b}{\rho ^{\Delta +}}$ where $\Delta _{\pm }$
are the roots of $\Delta \left( 3-\Delta \right) +m^{2}l^{2}=0$. When $-%
\frac{9}{4l^{2}}\leq m^{2}<-\frac{5}{4l^{2}}$ both branches are normalizable
but the $a$ branch contribute to the surface charges of the system. From the
form of the potential is possible to see that the mass is $m^{2}=-\frac{2}{%
l^{2}}$. When the mass is exactly such that $\frac{\Delta _{+}}{\Delta _{-}}%
=2$, the scalar field develops a logarithmic branch that, again, have a
non-trivial contribution to the charges at infinity. However this
logarithmic branch only appears if the expansion of the potential contains a
qubic term. Indeed, is possible to verify that with the change of
coordinates $r=\eta \exp (\frac{1}{\eta \rho }-\frac{1}{2\rho ^{2}\eta ^{2}}-%
\frac{\nu ^{2}-9}{24\eta ^{3}\rho ^{3}}),$ the scalar field takes the form

\begin{equation}
\phi =l_{\nu }^{-1}\left( \frac{1}{\eta \rho }-\frac{1}{2\rho ^{2}\eta ^{2}}-%
\frac{\nu ^{2}-9}{24\eta ^{3}\rho ^{3}}\right) ,
\end{equation}%
and the departure from the the AdS metric, defined by $ds^{2}=-(k+\frac{r^{2}%
}{l^{2}})dt^{2}+\frac{d\rho ^{2}}{(k+\frac{\rho ^{2}}{l^{2}})}+\rho
^{2}d\Sigma ,$ is

\begin{equation}
h_{mn}=\frac{\nu^{2}-4}{6\eta^{3}\rho}g_{mn}+O(\rho^{-2}),\qquad h_{tt}=%
\frac{\Lambda(\nu^{2}-4)}{18\eta^{3}\rho}+\frac{k\left( \nu-1\right)
+6M\nu^{2}\eta^{\nu}}{3\eta\rho}+O(\rho^{-2}),
\end{equation}

\begin{equation}
h_{\rho \rho }=\frac{3\left( \nu ^{2}-1\right) }{4\eta ^{2}\Lambda \rho ^{4}}%
+O(\rho ^{-5}),
\end{equation}%
where $g_{mn}$ are the components along $d\Sigma $. This coincides exactly
with eq. (6.2) of \cite{Henneaux:2006hk} with $\Delta _{-}=\Delta =1$, $a=%
\frac{1}{\eta l_{\nu }}$ and $b=-\frac{1}{2\eta ^{2}l_{\nu }}$. The case $%
\nu ^{2}=4 $ is peculiar in the sense that the deformation of the metric at
infinity is subleading than for generic $\nu $.

The cases with $\nu =2$ and $\nu =\infty $ are special, and can be treated
by a simple limiting procedure.

\subsection{$\protect\nu=2$}

Indeed, the potential (\ref{pot1}) has a smooth limit when $\nu =2,\,$which
is given by

\begin{equation}
V(\phi )=\frac{\alpha }{16\kappa }\left( \sinh (3\phi l_{2})+9\sinh (\phi
l_{2})-12\phi l_{2}\cosh \left( \phi l_{2}\right) \right) +\frac{\Lambda }{%
2\kappa \nu ^{2}}\left( e^{\phi l_{2}}+e^{-\phi l_{2}}\right) .  \label{pot3}
\end{equation}%
where $l_{2}=\sqrt{\frac{2\kappa }{3}}$. The metric functions also have a
smooth limit

\begin{equation}
\Omega(r)=\frac{4\eta r}{\left( r^{2}-\eta^{2}\right) ^{2}},
\end{equation}

\begin{equation}
F(r)=\frac{\eta ^{-2}\left( r^{2}-\eta ^{2}\right) ^{2}}{4}k+\left( \frac{3}{%
16}+\left( \frac{r}{2\eta }\right) ^{4}-\left( \frac{r}{2\eta }\right) ^{2}+%
\frac{1}{4}\ln (\frac{r}{\eta })\right) \alpha -\frac{\Lambda }{3}.
\end{equation}%
The potential (\ref{pot1}) has been considered in the context of the
existence of topological AdS black holes in \cite{Kolyvaris:2009pc}. When $%
\alpha =0$ and $k=-1$ this is the MTZ black hole \cite{Martinez:2004nb}.

\subsection{$\protect\nu=\infty$}

The $\nu =\infty $ case is a bit more subtle. First, is necessary to rescale
the area of the unit sphere as $d\Sigma \longrightarrow \nu ^{-2}d\Sigma $
which imply that the metric function $F$ rescale accordingly%
\begin{equation}
F(r)=r^{2-\nu }\eta ^{-\nu }\left( r^{\nu }-\eta ^{\nu }\right) ^{2}k+\left( 
\frac{1}{\left( \nu ^{2}-4\right) }-\left( 1+\frac{\eta ^{\nu }r^{-\nu }}{%
\nu -2}-\frac{\eta ^{-\nu }r^{\nu }}{\nu +2}\right) \frac{r^{2}}{\eta
^{2}\nu ^{2}}\right) \alpha ,
\end{equation}%
and the solution is now

\begin{equation}
ds^{2}=\Omega (r)(-F(r)dt^{2}+\frac{dr^{2}}{F(r)}+\nu ^{-2}d\Sigma ^{2}),
\end{equation}%
\begin{equation}
\Omega (r)=\frac{\nu ^{2}\eta ^{\nu -1}r^{\nu -1}}{\left( r^{\nu }-\eta
^{\nu }\right) ^{2}},\qquad \phi =l_{\nu }^{-1}\ln (r\eta ^{-1}).
\end{equation}

Let us introduce the changes of coordinates $r=\rho ^{\frac{1}{\nu }}$, $t=%
\frac{\tau }{\nu }$, and the reparameterization $\eta \rightarrow \eta ^{%
\frac{1}{\nu }}$, $\alpha \longrightarrow \nu ^{3}\alpha .$ The $\nu =\infty 
$ limit is then easily seen to give

\begin{equation}
ds^{2}=\Omega _{\infty }(\rho )(-F_{\infty }(\rho )d\tau ^{2}+\frac{d\rho
^{2}}{F_{\infty }(\rho )}+d\Sigma ^{2}),
\end{equation}%
\begin{equation}
\Omega _{\infty }(\rho )=\frac{\eta \rho }{\left( \rho -\eta \right) ^{2}}%
,\qquad \phi =\frac{1}{\sqrt{2\kappa }}\ln (\rho \eta ^{-1}),
\end{equation}

\begin{equation}
F_{\infty }(\rho )=\rho ^{-1}\eta ^{-1}\left( \rho -\eta \right)
^{2}k+\left( 2\ln (\frac{\eta }{\rho })+\frac{\rho }{\eta }-\frac{\eta }{%
\rho }\right) \alpha -\frac{\Lambda }{3},
\end{equation}

\begin{equation}
V_{\infty }(\phi )=\frac{2\alpha }{\kappa }\left( 2\phi l_{P}+\phi
l_{P}\cosh \left( \phi l_{P}\right) -3\sinh \left( \phi l_{P}\right) \right)
+\frac{\Lambda }{3}\left( 4+2\cosh (l_{P}\phi )\right)  \label{poti}
\end{equation}%
where $l_{P}=\sqrt{2\kappa }$ is proportional to the Planck length. The
potential (\ref{poti}) has been considered in the context of de Sitter black
hairy black holes compatible with inflation in \cite{Zloshchastiev:2004ny}.

\section*{Acknowledgments}

The author would like to thank the organizers of the conference
\textquotedblleft \textit{Relativity and Gravitation: 100 years after
Einstein in Prague\textquotedblright } for its excellent environment and
organization. Research of A.A. is supported in part by the Conicyt grant
Anillo ACT-91: \textquotedblleft Southern Theoretical Physics
Laboratory\textquotedblright\ (STPLab) and by the FONDECYT grant 11121187.

\end{document}